\documentclass{pos}

\title{Exploring the epsilon regime with lattice Wilson fermions}

\ShortTitle{Epsilon regime with Wilson fermions}

\author{Oliver B\"ar$^a$, \speaker{Silvia Necco}$^b$, Stefan
  Schaefer$^a$\\
$^a$  Institute of Physics, Humboldt University Berlin\\
Newtonstrasse 15, 12489 Berlin, Germany\\
$^b$  CERN, Physics Departement, 1211 Geneva 23, Switzerland \\

\vspace{0.1cm}
E-mail: \email{obaer@physik.hu-berlin.de, Silvia.Necco@cern.ch, sschaef@physik.hu-berlin.de}
}

\abstract{We study the impact of explicit chiral symmetry breaking of lattice Wilson fermions
on mesonic correlators in the $\epsilon$-regime using Wilson chiral perturbation theory.
We generalize the $\epsilon$-expansion of continuum chiral perturbation theory to nonzero lattice spacing $a$  and distinguish various regimes.
It turnes out that lattice corrections are highly suppressed, as long as quark masses are of the order $a\Lambda^2_{\rm QCD}$. The lattice spacing effects become more pronounced for smaller quark masses and may lead to non-trivial corrections of the continuum results at next-to-leading order. We compute these corrections for standard current and density correlation functions. A fit to lattice data shows that these corrections are small, as expected. 

\vspace{1.5cm}
\begin{flushright}
CERN-PH-TH/2009-052\\
SFB/CPP-09-34\\
HU-EP-09/16
\end{flushright}
         }

\FullConference{International Workshop on Effective Field Theories: from the pion to the upsilon \\
		February 2-6 2009\\
		Valencia, Spain}

\begin{document}

\section{Introduction}
Lattice QCD simulations with light quarks are now approaching domains where a reliable matching with the chiral effective theory can be performed. 
Through the matching it is possible to extract the Low-Energy Couplings (LECs) of the effective theory: many results have been presented in the past months for the leading order (LO) and next-to-leading order (NLO) couplings, both for the $N_f=2$ and $N_f=3$ effective theory (for a recent review see for instance \cite{Necco:2009cq}). 
A very important issue which concerns these determinations is the control over the systematic uncertainties. 
From this point of view, it is very useful to extract the LECs from a large set of observables and from different kinematic regimes:
this will give solidity to lattice results and will help to get a comprehensive picture of low energy properties of QCD.

An interesting approach is to investigate QCD in a finite volume $V=L^3 T$ in the so-called $\epsilon$-regime \cite{Gasser:1986vb,Gasser:1987ah}, where the pion wavelength is larger than the size of the box, $M_\pi L< 1$. The relevant feature of this regime is that, due to a reorganization of the chiral expansion,  volume effects are enhanced, 
while mass-effects are suppressed with respect to the usual infinite-volume case (or $p$-regime, where $M_\pi L\gg 1$).  For this reason, at a given order in the perturbative expansion, less LECs will appear: predictions are less ``contaminated'' by higher order unknown couplings, making the $\epsilon$-regime potentially  convenient for the extraction of the LO constants. 

From the point of view of lattice computations, reaching the $\epsilon$-regime requires simulating small quark masses; this may influence the choice of the discretized Dirac operator to be adopted.  Ideally, a Dirac operator which satisfies the Ginsparg-Wilson relation \cite{Ginsparg:1981bj} would be very advantageous, since it preserves chiral symmetry at finite lattice spacing, and small quark masses are accessible. Moreover, the topological charge can be unambiguously defined through the index theorem, and the lattice results may be matched with the predictions of the chiral effective theory at fixed topology \cite{Leutwyler:1992yt}.
A large number of quenched computations in the $\epsilon$-regime with Ginsparg-Wilson Dirac operator has been performed (see \cite{Giusti:2008fz} for a recent study, and references therein for precedent computations). Even though the results were to a large extend promising, the main obstacle for progress in real QCD is the fact that simulations with dynamical sea quarks are still extremely time-consuming. 
For recent calculations with dynamical chiral fermions see \cite{Fukaya:2007pn,DeGrand:2007tm,Hasenfratz:2007yj,Lang:2006ab}.

On the other hand, dynamical simulations with the Wilson Dirac operator with $O(a)$ improvement are becoming fairly inexpensive. 
Reaching small quark masses with Wilson fermions has been considered problematic for many years;  this issue can be faced by adopting techniques such as reweighting \cite{Hasenfratz:2008fg}.
This method has been applied in \cite{Hasenfratz:2008ce} allowing to reach the $\epsilon$-regime. Also the ETM collaboration investigated the $\epsilon$-regime with a twisted mass Wilson Dirac operator \cite{Jansen:2007rx}.

Since Wilson fermions explicitly break the chiral symmetry at finite lattice spacing, the matching with the chiral effective theory should be performed only after a continuum extrapolation of the lattice results. While this is not an unrealistic goal for the near future, the presently available simulations in the $\epsilon$-regime are carried out at a single value of the lattice spacing.  
In \cite{Hasenfratz:2008ce} the pseudoscalar and axial correlations functions turned out to be very well described by the predictions of the continuum chiral effective theory at NLO. Similar observations have been made by ETM. Still, it is important to have a theoretical understanding of the impact
of explicit breaking of chiral symmetry on computations in the $\epsilon$-regime.
We address this question in \cite{Bar:2008th}: 
the tool that we adopt is the so-called Wilson Chiral Perturbation Theory (WChPT) \cite{Sharpe:1998xm,Rupak:2002sm}, the low-energy effective theory for lattice QCD with Wilson Dirac operator. 
A similar analysis has been carried out in \cite{Shindler:2009ri}.

\section{Wilson Chiral Perturbation Theory}
The chiral effective Lagrangian of WChPT is expanded in powers of pion momenta $p^2$, the quark mass $m$ and lattice spacing $a$. Based on symmetries of the underlying Symanzik action \cite{Symanzik:1983dc}, the chiral Lagrangian including all terms of $O(p^4,p^2m,m^2,p^2a,ma)$ is given in \cite{Rupak:2002sm}. The $O(a^2)$ contributions are constructed in \cite{Bar:2003mh,Aoki:2003yv}.
In the following we will restrict ourselves to the case $N_f=2$ with degenerate quark mass $m$. 
The leading order Euclidean chiral Lagrangian in the continuum is given by \cite{Weinberg:1978kz,Gasser:1983yg}
\begin{equation}\label{eq1}
\mathcal{L}_2=\frac{F^2}{4}{\rm Tr}\left(\partial_\mu U\partial_\mu U^\dagger   \right)
-\frac{F^2Bm}{2}{\rm Tr}\left( U + U^\dagger\right).
\end{equation}
The pseudo Nambu-Goldstone modes are parametrized as usual by the SU(2) field 
$$
U(x)=\exp\left(2i\xi(x)\right)/F,
$$ 
and $F$, $B$ are the familiar LO couplings. The leading terms involving the lattice spacing are
\begin{eqnarray}
\mathcal{L}_a& = & \hat{a}W_{45}{\rm Tr}\left(\partial_\mu U\partial_\mu
    U^\dagger \right){\rm
    Tr}\left(U+U^\dagger\right)-\hat{a}\hat{m}W_{68}\left({\rm
      Tr}\left(U+U^\dagger\right)\right)^2,\\\label{eq2}
\mathcal{L}_{a^2}& = &\frac{F^2}{16}c_2 a^2 \left({\rm
      Tr}\left(U+U^\dagger\right)\right)^2,\label{eq3}
\end{eqnarray}
where $\hat{m}= 2Bm$ and $\hat{a}=2W_0 a$. $W_{45}$, $W_{68}$, $W_0$ and $c_2$ are new LECs which are not determined by the symmetries. Note that the mass parameter $m$ in Eq. (\ref{eq2}) is the so-called \emph{shifted} mass \cite{Sharpe:1998xm}: besides the dominant additive mass renormalization proportional to $1/a$ it also contains the leading correction of $O(a)$. 

Currents and densities in WChPT can be constructed by a standard spurion analysis or by introducing source terms. Here we report the axial vector current and the pseudoscalar density including the leading $O(a)$ corrections \cite{Sharpe:2004ny,Aoki:2007es}:
\begin{eqnarray}
A^a_{\mu,{\rm WChPT}} & = & A^a_{\mu,{\rm cont}} \left\{ 1+\frac{4}{F^2}\hat{a}\left[ W_{45}{\rm Tr}(U+U^\dagger)+4W_A  \right]  \right\}  +2\hat{a}W_{10}\partial_{\mu}
{\rm Tr}\left(T^a(U-U^\dagger)\right),     \label{eq3a}  \\ 
P^a_{\rm WChPT}& = &  P^a_{\rm cont}   \left\{ 1+\frac{4}{F^2}\hat{a}\left[ W_{68}{\rm Tr}(U+U^\dagger)+4W_P \right]  \right\},       \label{eq3b}
\end{eqnarray}
where 
\begin{equation}\label{eq3d}
A^a_{\mu,{\rm cont}}  =  i\frac{F^2}{2}{\rm Tr} \left(T^a(U^\dagger\partial_\mu U -U\partial_\mu U^\dagger)   \right),\;\;\;\;\;\;\;\;\;\;\;
P^a_{\rm cont}   =  i\frac{F^2B}{2} {\rm Tr}\left(T^a(U-U^\dagger)   \right),
\end{equation}
and $T^a$ are SU(2) generators normalized such that ${\rm Tr}(T^a T^b)=\delta^{ab}/2$. Notice that the LECs $W_{A,P}$ stem from the renormalization factors, which up to $O(a)$ take the form $Z_{A,P}=1+16\hat{a}W_{A,P}/F^2$. 

\subsection{Power counting in infinite volume}
In WChPT there are two parameters which break explicitly the chiral symmetry, the quark mass $m$ (counted as $O(p^2)$) and the lattice spacing $a$. 
The power counting is determined by the relative size of these two parameters. In particular, one distinguishes \cite{Sharpe:2004ny,Sharpe:2004ps} two different regimes: (i) the GSM \footnote{GSM stands for \emph{generically small masses}.} regime, where $m\sim a\Lambda^2_{\rm QCD}$ and (ii) the Aoki regime where $m\sim a^2\Lambda^3_{\rm QCD}$. In the Aoki regime lattice artefacts are more pronounced, and the $\mathcal{L}_{a^2}$ in Eq. (\ref{eq3}) enters already at LO. The pion mass at leading order is given by
\begin{eqnarray}
{\rm GSM\; regime} & : & M_0^2=2Bm,\\
{\rm Aoki\; regime} & : & M_0^2=2Bm-2c_2a^2. \label{maoki}
\end{eqnarray}
The sign of $c_2$ governs the phase diagram of the theory. The reader can refer to \cite{Sharpe:1998xm} for a complete discussion.

\subsection{Power counting in the $\epsilon$-regime}
As already anticipated in the introduction, the $\epsilon$-regime is a finite-volume regime where the pion wavelength is larger than the size of the box, $M_\pi L<1$ (but still $L\gg 1/\Lambda_{\rm QCD}$) \cite{Gasser:1986vb,Gasser:1987ah}. This corresponds to approaching the chiral limit by keeping the dimensionless quantity $\mu=m\Sigma V\lesssim O(1)$ (where $\Sigma=F^2B$ is the quark condensate in the chiral limit).  The main effect of formulating the effective theory in this regime is that the pion zero-mode becomes non-perturbative and its contribution has to be treated exactly.  This is achieved by factorizing the pseudo Nambu-Goldstone boson fields as
\begin{equation}
U(x)=\exp\left(\frac{2i}{F}\xi(x)   \right)U_0,
\end{equation}
where the constant $U_0\in$ SU(2) represents the collective zero-mode. The non-zero modes $\xi$ can be still treated perturbatively. 
The $\epsilon$-regime requires a reorganization of the perturbative series: in the continuum, this corresponds to taking the quark mass of order $m\sim O(\epsilon^4)$. 
Mass effects are hence suppressed compared to the $p$-regime (or infinite volume) case, while finite-volume effects are enhanced and become polynomial in $(FL)^{-2}$. 

We now want to extend the WChPT to the $ \epsilon$-regime.  Also in this case we have to assign a relative power counting of the lattice spacing $a$ with respect to the quark mass $m$. If we assume that the quark mass can be considered of order $m\sim O(\epsilon^4)$ also in  WChPT 
\footnote{While this is a natural choice in the GSM regime, the situation in the Aoki regime can be more subtle. See \protect\cite{Bar:2008th} for a more detailed discussion on this subject. }, we obtain 
\begin{eqnarray}
{\rm{GSM\; regime}} &: & m\sim O(a\Lambda^2_{\rm QCD})\;\rightarrow\; a\sim O(\epsilon^4),     \\
{\rm{Aoki\; regime}} &: & m\sim O(a^2\Lambda^3_{\rm QCD})\;\rightarrow\; a\sim O(\epsilon^2).   
\end{eqnarray}
Moreover, the $\epsilon$- expansion allows us to introduce another intermediate regime between the GSM and the Aoki regime: we can define the GSM$^*$ regime, where $a\sim O(\epsilon^3)$. 

We are interested in computing two-point correlation functions within the WChPT in the $\epsilon$-regime. In particular, we give explicit results for the pseudoscalar and axial time correlators, 
\begin{equation}
\delta^{ab}C_{PP}(t)=\int d^3\vec{x}\langle P^a(x) P^b(0)\rangle,\;\;\;\; \delta^{ab}C_{AA}(t)=\int d^3\vec{x}\langle A_0^a(x) A_0^b(0)\rangle.
\end{equation}
Currents and densities are defined in Eqs. (\ref{eq3a}, \ref{eq3b}); the subscript ``WChPT'' is now omitted.
By adopting the power counting that we have defined, we find that in the GSM regime lattice corrections enter only at NNLO, while in the  GSM$^*$ regime they appear already at NLO. In particular, the leading correction is given only by the term $\mathcal{L}_{a^2}$ in Eq. (\ref{eq3}).
 The reason of this suppression can be traced back to the fact that the lattice spacing corrections in the chiral effective theory action and in the effective operators are either quadratic in $a$ or they come with an additional power of either $m$ or $p^2$. Hence, in the $\epsilon$-regime the suppression of lattice spacing corrections works similarly as the suppression of mass effects.
On the other hand, in the Aoki regime effects of lattice artefacts are more severe and show up already at LO. 

It is important to observe that these considerations are valid for unimproved Wilson fermions.  If the theory is non-perturbatively $O(a)$ improved, the corrections 
due to $\mathcal{L}_{a}$ as well as the $O(a)$ terms in the operators are absent, and lattice artefacts are due to $\mathcal{L}_{a^2}$ only. 
Consequently, in the $\epsilon$-regime the leading corrections due to lattice artefacts are essentially unaltered for the unimproved theory, since improvement acts only on subleading terms. 
\subsection{Leading corrections in the GSM$^*$ regime}
The continuum pseudoscalar and axial correlators at NLO in the $\epsilon$-expansion can be written as \cite{Hansen:1990un}
\begin{equation}\label{eq5}
C_{PP,AA\;\rm{ct}}(t)=a_{P,A} + b_{P,A} h_1(t/T),
\end{equation}
where
\begin{equation}
h_1(\tau)=\frac{1}{2}\left[\left(| \tau|-\frac{1}{2}  \right)^2 -\frac{1}{12} \right].
\end{equation}
For $N_f=2$ the coefficients $a_{P,A},b_{P,A}$ explicitly read \cite{Hansen:1990un}
\begin{eqnarray}
a_P & = & \frac{L^3}{2}\frac{\Sigma^2_{\rm eff}}{\mu_{\rm eff}} \frac{I_2(2\mu_{\rm eff})}{I_1(2\mu_{\rm eff})},\;\;\;\;\;\;\;\;\;\;\;\;\;\;\;\;\;\; b_P =  \frac{T\Sigma^2}{2F^2} \left[2-\frac{1}{\mu}\frac{I_2(2\mu)}{I_1(2\mu)}   \right],                      \label{eq7}\\
a_A & = & -\frac{F^2}{T} \left[1-\frac{I_2(2\mu_{\rm eff})}{\mu_{\rm eff}I_1(2\mu_{\rm eff}) }    \right]  -\frac{2\beta_1}{T\sqrt{V}}  \left[2-\frac{1}{\mu}\frac{I_2(2\mu)}{I_1(2\mu)}   \right]      +\frac{2T}{V}k_{00}\frac{I_2(2\mu)}{\mu I_1(2\mu)},    \nonumber  \\
b_A & = &  -\frac{2T}{V}  \frac{\mu I_2(2\mu)}{I_1(2\mu)}.  \label{eq7a}
\end{eqnarray}
$I_1$, $I_2$ are modified Bessel functions of the first kind; $\beta_1$ and $k_{00}$ are so-called shape factors \cite{Hasenfratz:1989pk,Hansen:1990un}, which depend only on the geometry of the finite box.
$\Sigma_{\rm eff}$ is the quark condensate at one loop \cite{Gasser:1987ah}
\begin{equation}
\Sigma_{\rm eff}=\Sigma\left( 1+\frac{3}{2F^2}\frac{\beta_1}{\sqrt{V}}  \right),
\end{equation}
and $\mu_{\rm eff}=m\Sigma_{\rm eff} V$. As already anticipated, the continuum NLO predictions in the $\epsilon$-regime contain only the LO LECs $\Sigma$ and $F$.

The first non-trivial modification of the continuum NLO results appear in the GSM$^*$ regime and it is due to $\mathcal{L}_{a^2}$ only. 
In this case we can write down the full NLO correlators in WChPT as
\begin{equation}
C_{PP,AA}(t)= C_{PP,AA\;\rm{ct}}(t)+ C_{PP,AA\; a^2}(t).
\end{equation}
By performing the explicit computation (see \cite{Bar:2008th} for the full details) it turns out that the corrections $C_{PP,AA\; a^2}(t)$ are time-independent and hence affect only the constant part of the correlators. In particular we obtain:
\begin{equation}
C_{PP a^2}(t)= \frac{L^3\Sigma^2}{2}\rho\Delta_{a^2},\;\;\;\;C_{AA a^2}(t) = \frac{F^2}{T}\rho\Delta_{a^2},
\end{equation}
where
\begin{equation}\label{eq9}
\Delta_{a^2}=\frac{5\mu I_1^2(2\mu)-10I_1(2\mu)I_2(2\mu)-3\mu I_2^2(2\mu)}{2\mu^3I_1^2(2\mu)},
\end{equation}
and $\rho=F^2c_2a^2V$ is the dimensionless LEC which parametrizes the $O(a^2)$ correction.

It is useful to compute the leading $O(a^2)$ corrections to the PCAC quark mass:
\begin{equation}
m_{\rm PCAC}=m\left[1+\rho \left(\frac{2}{\mu^2} -\frac{I_1(2\mu)}{\mu I_2(2\mu)}   \right)    \right].
\end{equation}
It is now possible to express the correlators $C_{PP,AA}(t)$ as a function of $\tilde{\mu}=m_{PCAC}\Sigma V$; the result is
\begin{equation}
C_{PP}(t)= C_{PP\rm{ct}}(t)+ \frac{L^3\Sigma^2}{2}\rho\tilde\Delta_{a^2}, \;\;\;\;\;C_{AA}(t)= C_{AA\rm{ct}}(t)+\frac{F^2}{T}\rho\tilde\Delta_{a^2},
\end{equation}
where the continuum correlators are as in Eq. (\ref{eq5}), but with the replacements $\mu\rightarrow \tilde\mu$, $\mu_{\rm eff}\rightarrow \tilde\mu_{\rm eff}=m_{\rm PCAC}\Sigma_{\rm eff} V$ and
\begin{equation}
\tilde\Delta_{a^2}=\frac{4\tilde\mu^2I_1^3(2\tilde\mu)-11\tilde\mu I_1^2(2\tilde\mu)I_2(2\tilde\mu)+2(3-2\tilde\mu^2)I_1(2\tilde\mu)I_2^2(2\tilde\mu)+5\tilde\mu I_2^3(2\tilde\mu)     }{2\tilde\mu^3I_1^2(2\tilde\mu)I_2(2\tilde\mu)}.
\end{equation}
Other correlation functions can be computed along the same line. For instance, in  \cite{Bar:2008th} we report also the result for the vector correlator.  
\section{Reanalysis of lattice data and conclusions}
\begin{figure}
\begin{minipage}{7.8cm}
\includegraphics[width=8.3cm]{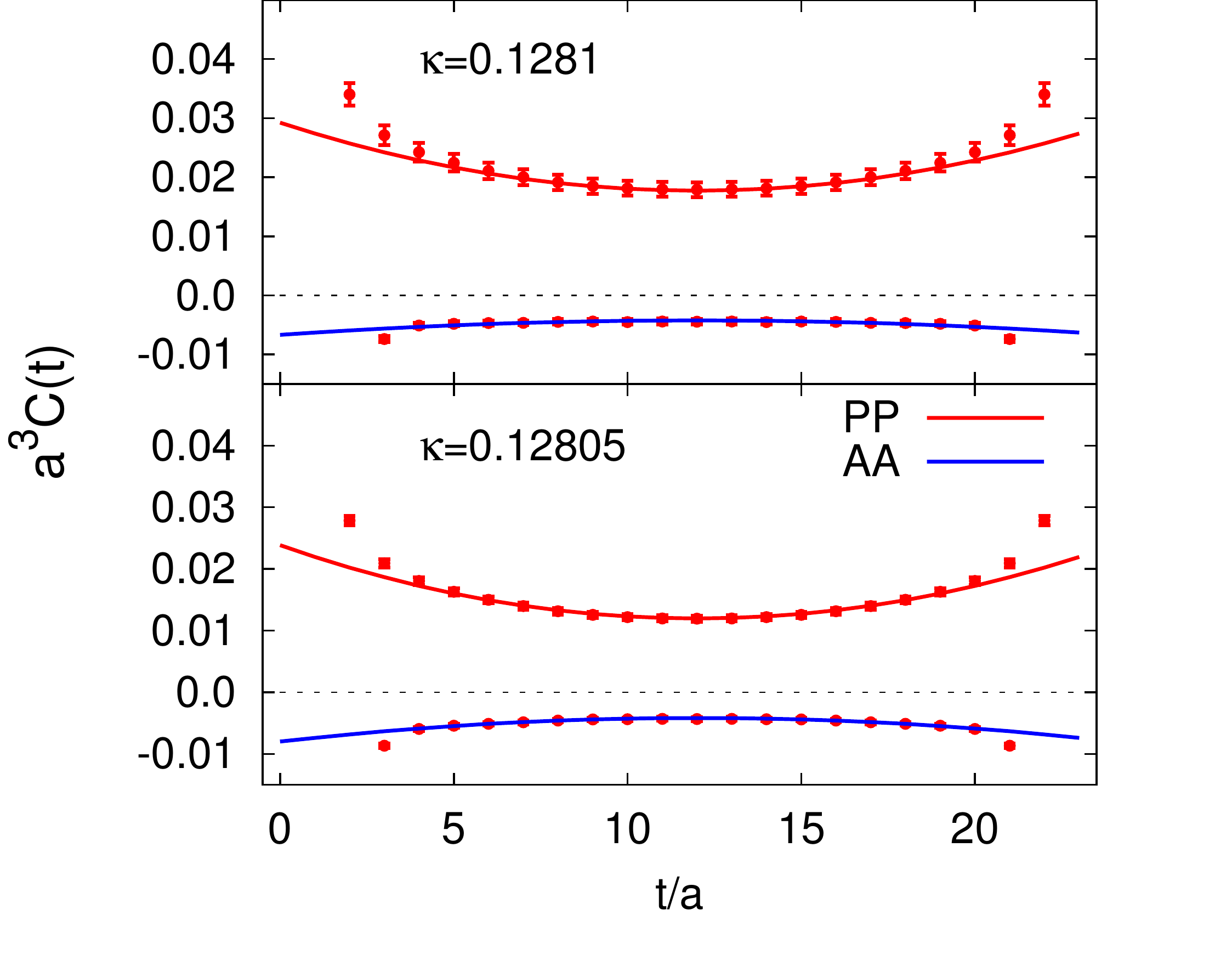}
\end{minipage}
\begin{minipage}{7.8cm}
\includegraphics[width=8.3cm]{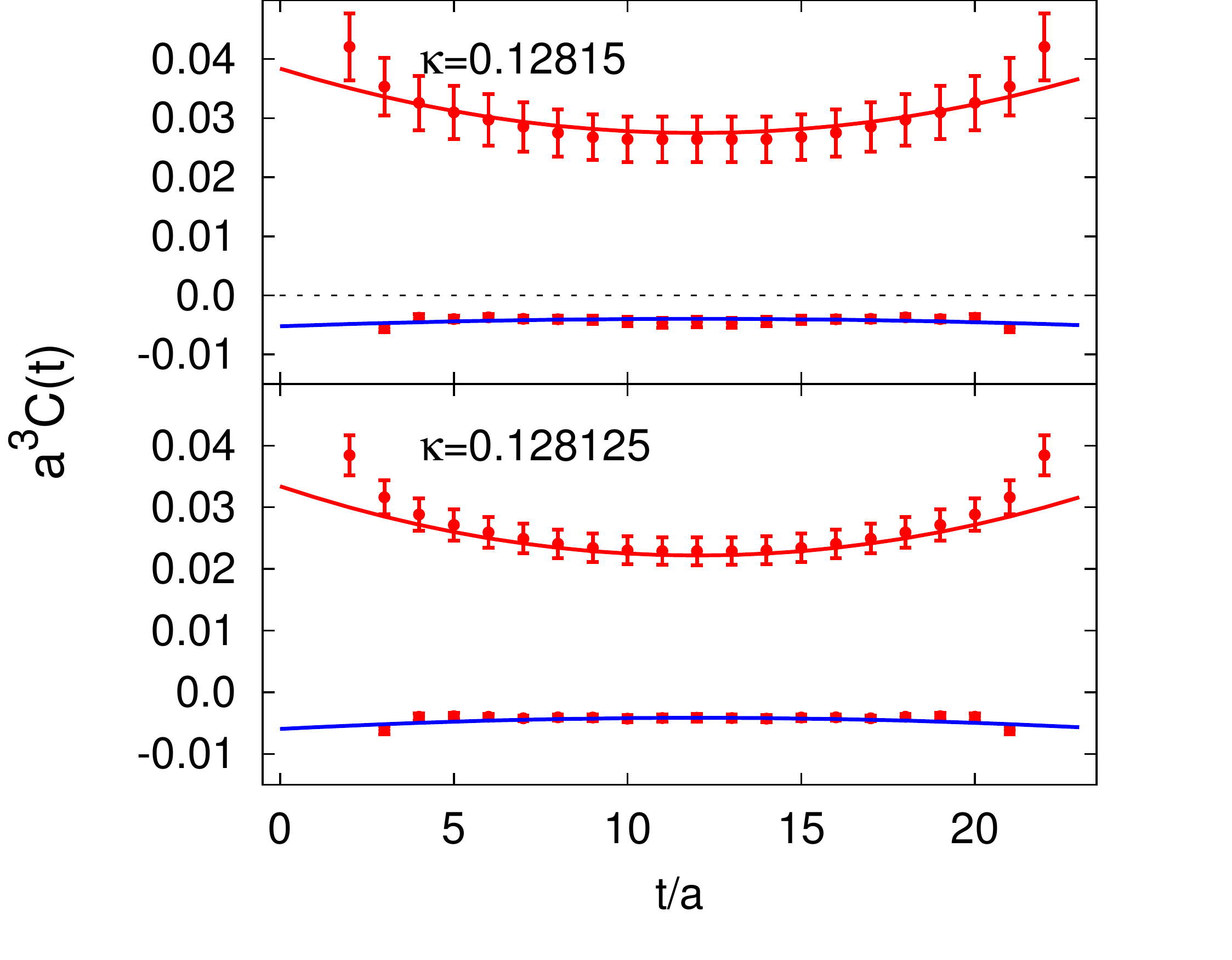}
\end{minipage}
\caption{Fit of the WChPT predictions to lattice data. All data points within the fit range of $t/a\in[6,18]$ for the four sea quark masses are included in the combined fit. The hopping parameter $\kappa=(0.128150,0.128125,0.1281,0.128050)$ corresponds to $am_{\rm PCAC}=(0.0019(4),0.0024(3),0.0030(3),0.0044(3))$ respectively.
The axial vector correlator is multiplied by a factor 50 for better visibility.  }\label{fig1}
\end{figure}

These predictions from WChPT at NLO can be tested against lattice data generated in \cite{Hasenfratz:2008ce}, where pseudoscalar and axial correlators have been computed on an ensemble with $N_f=2$ flavours of dynamical improved NHYP Wilson fermions \cite{Hasenfratz:2007rf}. 
The lattice spacing is $a\simeq 0.115$ fm, and two lattice extents are available, 
$L_1=16a\simeq 1.84$ fm and $L_2=24a\simeq 2.8$ fm. Quark masses approach the $\epsilon$-regime, with $\tilde\mu\simeq 0.7-2.9$ for the volume $V_1=L_1^4$ and  $\tilde\mu\simeq 2.1-5.0$ for the volume $V_2=L_2^4$.
In the GSM$^*$ regime, we have only the additional LEC $c_2$ with respect to the continuum case. Notice that its value will depend on the particular discretized action which is used. 
We simultaneously fit the two correlators for all      available quark masses; for the volume $V_2$, a fit in the range $t\in[6,18]$ gives
\begin{equation}\label{res}
\left[\Sigma^{\overline{\rm MS}}(\mu=2\;{\rm GeV})\right]^{1/3} =249(4)\; {\rm MeV},\;\;\; F=88(3) \; {\rm MeV},\;\;\;c_2=0.02(8)\;{\rm GeV}^4.
\end{equation}
The data, along with the theoretical curves, are shown in Fig. \ref{fig1}. The errors from the renormalization factors $Z_A$, $Z^{\overline{\rm MS}}_P(\mu=2\;{\rm GeV})$ computed in \cite{Hasenfratz:2008ce} are not included in the uncertainties of the LECs. Varying the time range of the fit does not give significant differences for the LECs, as long as $t_{\rm min}/a >4$. Also discarding the heaviest mass does not change the results of Eq. (\ref{res}) within the statistical errors. 
The smallest volume $V_1$ yields values which are consistent with Eq. (\ref{res}), but the large $\chi^2$ of the fit may indicate that NLO formulae are no longer applicable. 
The values of $F$ and $\Sigma$ are compatible with other determinations \cite{Necco:2009cq}, while the value of $c_2$ is compatible with zero. 
 A continuum fit (with $c_2=0$) yields virtually unchanged values for $F$ and $\Sigma$, showing that cut-off effects do not impact the extraction of the LECs beyond the level of the statistical uncertainties.

This is a very encouraging result: simulations with Wilson fermions in the $\epsilon$-regime are feasible 
and seem to be a viable alternative to dynamical simulations with chiral fermions. Similar conclusions have been reached in \cite{Shindler:2009ri}.
The results derived here can be generalized in various ways, for example to the case with a twisted mass term or to an arbitrary number of flavors. 

\providecommand{\href}[2]{#2}\begingroup\raggedright\endgroup


\begin{thebibliography}{10}

\bibitem{Necco:2009cq}
S.~Necco, 
  \href{http://xxx.lanl.gov/abs/0901.4257}{{\tt arXiv:0901.4257}}.

\bibitem{Gasser:1986vb}
J.~Gasser and H.~Leutwyler, {\em Phys.
  Lett.} {\bf B184} (1987) 83.

\bibitem{Gasser:1987ah}
J.~Gasser and H.~Leutwyler, {\em
  Phys. Lett.} {\bf B188} (1987) 477.

\bibitem{Ginsparg:1981bj}
P.~H. Ginsparg and K.~G. Wilson, {\em Phys. Rev.} {\bf D25} (1982) 2649.

\bibitem{Leutwyler:1992yt}
H.~Leutwyler and A.~Smilga,   {\em Phys. Rev.} {\bf D46} (1992) 5607--5632.

\bibitem{Giusti:2008fz}
L.~Giusti {\em et.~al.},  {\em JHEP} {\bf 05} (2008) 024,
  [\href{http://xxx.lanl.gov/abs/0803.2772}{{\tt arXiv:0803.2772}}].


\bibitem{Fukaya:2007pn}
{\bf JLQCD} Collaboration, H.~Fukaya {\em et.~al.},   {\em Phys.
  Rev.} {\bf D77} (2008) 074503, [\href{http://xxx.lanl.gov/abs/0711.4965}{{\tt
  arXiv:0711.4965}}].

\bibitem{DeGrand:2007tm}
T.~DeGrand and S.~Schaefer,  {\em Phys. Rev.} {\bf D76} (2007)
  094509, [\href{http://xxx.lanl.gov/abs/0708.1731}{{\tt arXiv:0708.1731}}].

\bibitem{Hasenfratz:2007yj}
P.~Hasenfratz {\em et.~al.}q,
  \href{http://xxx.lanl.gov/abs/0707.0071}{{\tt 0707.0071}}.

\bibitem{Lang:2006ab}
C.~B. Lang, P.~Majumdar, and W.~Ortner,  {\em Phys. Lett.} {\bf B649} (2007)
  225--229, [\href{http://xxx.lanl.gov/abs/hep-lat/0611010}{{\tt
  hep-lat/0611010}}].

\bibitem{Hasenfratz:2008fg}
A.~Hasenfratz, R.~Hoffmann, and S.~Schaefer,  {\em Phys. Rev.} {\bf D78} (2008) 014515,
  [\href{http://xxx.lanl.gov/abs/0805.2369}{{\tt arXiv:0805.2369}}].

\bibitem{Hasenfratz:2008ce}
A.~Hasenfratz, R.~Hoffmann, and S.~Schaefer,  {\em Phys.
  Rev.} {\bf D78} (2008) 054511, [\href{http://xxx.lanl.gov/abs/0806.4586}{{\tt
  arXiv:0806.4586}}].

\bibitem{Jansen:2007rx}
K.~Jansen, A.~Nube, A.~Shindler, C.~Urbach, and U.~Wenger,  {\em PoS} {\bf LAT2007} (2007)
  084, [\href{http://xxx.lanl.gov/abs/0711.1871}{{\tt arXiv:0711.1871}}];
K.~Jansen, A.~Nube, and A.~Shindler, \href{http://xxx.lanl.gov/abs/0810.0300}{{\tt
  arXiv:0810.0300}}.

\bibitem{Bar:2008th}
O.~B\"ar, S.~Necco, and S.~Schaefer,   {\em JHEP} {\bf 03} (2009) 006,
  [\href{http://xxx.lanl.gov/abs/0812.2403}{{\tt arXiv:0812.2403}}].

\bibitem{Sharpe:1998xm}
S.~R. Sharpe and J.~Singleton, Robert~L.,  {\em Phys. Rev.} {\bf D58} (1998) 074501,
  [\href{http://xxx.lanl.gov/abs/hep-lat/9804028}{{\tt hep-lat/9804028}}].

\bibitem{Rupak:2002sm}
G.~Rupak and N.~Shoresh, {\em Phys. Rev.} {\bf D66} (2002) 054503,
  [\href{http://xxx.lanl.gov/abs/hep-lat/0201019}{{\tt hep-lat/0201019}}].

\bibitem{Shindler:2009ri}
A.~Shindler, 
   {\em Phys. Lett.} {\bf B672} (2009) 82--88,
  [\href{http://xxx.lanl.gov/abs/0812.2251}{{\tt arXiv:0812.2251}}].

\bibitem{Symanzik:1983dc}
K.~Symanzik,   {\em Nucl. Phys.} {\bf B226} (1983) 187; 
{\em Nucl. Phys.} {\bf
  B226} (1983) 205.

\bibitem{Bar:2003mh}
O.~B\"ar, G.~Rupak, and N.~Shoresh,   {\em Phys. Rev.} {\bf D70} (2004) 034508,
  [\href{http://xxx.lanl.gov/abs/hep-lat/0306021}{{\tt hep-lat/0306021}}].

\bibitem{Aoki:2003yv}
S.~Aoki,  {\em Phys. Rev.} {\bf D68} (2003)
  054508, [\href{http://xxx.lanl.gov/abs/hep-lat/0306027}{{\tt
  hep-lat/0306027}}].

\bibitem{Weinberg:1978kz}
S.~Weinberg, {\em Physica} {\bf A96}
  (1979) 327.

\bibitem{Gasser:1983yg}
J.~Gasser and H.~Leutwyler,  {\em
  Ann. Phys.} {\bf 158} (1984) 142.

\bibitem{Sharpe:2004ny}
S.~R. Sharpe and J.~M.~S. Wu,   {\em Phys. Rev.} {\bf D71} (2005) 074501,
  [\href{http://xxx.lanl.gov/abs/hep-lat/0411021}{{\tt hep-lat/0411021}}].

\bibitem{Aoki:2007es}
S.~Aoki and O.~B\"ar, 
   {\em PoS} {\bf LAT2007} (2007) 062,
  [\href{http://xxx.lanl.gov/abs/0710.0072}{{\tt arXiv:0710.0072}}].

\bibitem{Sharpe:2004ps}
S.~R. Sharpe and J.~M.~S. Wu,   {\em Phys. Rev.} {\bf D70} (2004) 094029,
  [\href{http://xxx.lanl.gov/abs/hep-lat/0407025}{{\tt hep-lat/0407025}}].

\bibitem{Hansen:1990un}
F.~C. Hansen,   {\em Nucl. Phys.} {\bf B345} (1990) 685--708.

\bibitem{Hasenfratz:1989pk}
P.~Hasenfratz and H.~Leutwyler,   {\em
  Nucl. Phys.} {\bf B343} (1990) 241--284.

\bibitem{Hasenfratz:2007rf}
A.~Hasenfratz, R.~Hoffmann, and S.~Schaefer,   {\em JHEP} {\bf 05} (2007) 029,
  [\href{http://xxx.lanl.gov/abs/hep-lat/0702028}{{\tt hep-lat/0702028}}].

\end{thebibliography}
\end{document}